\documentclass[preprint,aps,tightenlines,draft]{revtex4}

\usepackage{amsmath,amssymb,psfig,epsf}

\begin{document}
\def\eqn#1{Eq.$\,$#1}
\def\mb#1{\setbox0=\hbox{$#1$}\kern-.025em\copy0\kern-\wd0
\kern-0.05em\copy0\kern-\wd0\kern-.025em\raise.0233em\box0}
\preprint{}

\title{On the interpretations of Tsallis functional in
connection with Vlasov-Poisson and related systems: 
Dynamics vs Thermodynamics}
\author{P.H. Chavanis and C. Sire }
\affiliation{Laboratoire de Physique Th\'eorique (UMR 5152 du CNRS),
Universit\'e Paul Sabatier, 118 route de Narbonne, 31062 Toulouse
Cedex 4, France}

\begin{abstract}

\vskip 0.5cm

We discuss different interpretations of Tsallis functional in
astrophysics. In principle, for $t\rightarrow +\infty$, a
self-gravitating system should reach a statistical equilibrium state
described by the Boltzmann distribution. However, this tendency is
hampered by the escape of stars and the gravothermal
catastrophe. Furthermore, the relaxation time increases almost
linearly with the number of particles $N$ so that most stellar systems
are in a collisionless regime described by the Vlasov equation. This
equation admits an infinite number of stationary solutions. The system
can be trapped in one of them as a result of phase mixing and
incomplete violent relaxation and remains frozen in this
quasi-stationary state for a very long time until collisional effects
finally come into play. Tsallis distribution functions form a
particular class of stationary solutions of the Vlasov equation named
stellar polytropes. We interpret Tsallis functional as a particular
H-function in the sense of Tremaine, H\'enon and Lynden-Bell
[Mon. Not. R. astr. Soc.  {\bf 219}, 285 (1986)]. Furthermore, we show
that the criterion of nonlinear dynamical stability for spherical
stellar systems described by the Vlasov-Poisson system resembles a
criterion of thermodynamical stability in the microcanonical ensemble
and that the criterion of nonlinear dynamical stability for barotropic
stars described by the Euler-Poisson system resembles a criterion of
thermodynamical stability in the canonical ensemble.  Accordingly, a
{\it thermodynamical analogy} can be developed to investigate the
nonlinear dynamical stability of barotropic stars and spherical
galaxies but the notions of entropy, free energy and temperature are
essentially effective. This analogy provides an interpretation of the
nonlinear Antonov first law in terms of ensemble
inequivalence. Similar ideas apply to other systems with long-range
interactions such as two-dimensional vortices and the HMF model. We
propose a general scenario to understand the emergence of coherent
structures in long-range systems and discuss the
dynamical/thermodynamical ``duality'' of their description. We stress
that the thermodynamical analogy that we develop is only valid for
systems whose distribution function depends only on energy. We discuss
two other, independent, interpretations of Tsallis functional in
astrophysics, in relation with generalized kinetic equations and
quasi-equilibrium states of collisional stellar systems.

\pacs{???}
\vskip0.5cm
\noindent Corresponding author:
P.H. Chavanis; e-mail: chavanis@irsamc.ups-tlse.fr; Tel:
+33-5-61558231; Fax: +33-5-61556065

\end{abstract}

\maketitle

\newpage

\section{Introduction}
\label{sec_introduction}

For $t\rightarrow +\infty$, a self-gravitating system is expected to
achieve a statistical equilibrium state described by the Boltzmann
distribution. This statistical approach is the correct description of
globular clusters, which are small groups of stars ($N\sim 10^{6}$)
whose age is of the same order as the relaxation time. Their observed
thermal distribution (Michie-King models) is the natural outcome of a
``collisional'' relaxation due to the development of stellar
encounters. This collisional relaxation is usually described by the
gravitational kinetic Landau equation which satisfies a H-theorem for
the Boltzmann entropy. Fundamentally, the Boltzmann entropy is equal
to the logarithm of disorder where the disorder measures the number of
microstates (complexions) associated with a given macrostate. The
maximization of the Boltzmann entropy at fixed mass and energy has a
clear statistical and thermodynamical meaning: it determines the most
probable distribution of stars at equilibrium, assuming that the
microstates are equiprobable. However, unlike ordinary systems, the
statistical mechanics of self-gravitating systems encounters some
difficulties due to the long-range unshielded nature of gravity and
the divergence of the potential at short distances
\cite{paddy,houches}. In particular, there is no statistical
equilibrium state in a strict sense: the Boltzmann entropy has no
global maximum and isothermal distributions have infinite mass
\cite{bt}. The increase of entropy as  the system spreads reflects
the natural tendency of stellar clusters to evaporate (this is
already the case for a classical gas, without gravity, if it is
not confined by the walls of a container). A statistical
equilibrium state can be defined only in a weaker sense if the
system is tidally truncated or confined by the pressure exerted
by an external medium. Theorists overcome the infinite mass
problem by confining the system within a box. Physically, the box
delimits the region of space where the system can be assumed
isolated and described by thermodynamical arguments. Typically,
the box radius plays the same role as the tidal radius. However,
even for box-confined systems, statistical equilibrium does not
exist if the energy is too low. Below the Antonov threshold
\cite{antonov}, the system undergoes gravitational collapse called
gravothermal catastrophe \cite{lbw}. This core collapse ultimately
leads to the formation of binaries \cite{henon}. Statistical
equilibrium states exist at sufficiently high energies. They are
metastable (local entropy maxima) but their lifetime is considerable
as it increases exponentially rapidly with the number of particles
\cite{meta}.  Therefore, small clusters of stars such as globular
clusters correspond to long-lived metastable states. Furthermore, for
self-gravitating systems, statistical ensembles are non-equivalent
\cite{paddy,aa3}  and, for isolated systems, only the microcanonical
ensemble makes sense. To our opinion, the evaporation of stellar
clusters, the gravothermal catastrophe, the formation of binaries and
the long-lived metastability of globular clusters are the correct
answers to the absence of strict statistical equilibrium for
self-gravitating systems. They correspond to important physical
processes that have been clearly identified in astrophysics. {\it With
these limitations in mind, self-gravitating systems can be described
by conventional statistical mechanics
\cite{paddy,houches}.}

However, for systems with long-range interactions, the collisional
relaxation time is extremely long because it increases algebraically
with the number $N$ of particles. For example, for stellar systems,
the Chandrasekhar relaxation time scales as $t_{relax}\sim (N/\ln
N)t_{D}$, where $t_{D}$ is the dynamical time. Therefore, for a very
long period, encounters between stars are negligible and the galactic
dynamics is described by the Vlasov, or collisionless Boltzmann,
equation coupled to gravity via the Poisson equation
\cite{bt,houches}. The Vlasov-Poisson system is the correct description of
large groups of stars such as elliptical galaxies ($N\sim 10^{12}$)
that are still, at the present time, in the collisionless regime
(contrary to globular clusters). Now, the Vlasov equation admits an
infinite class of stationary solutions.  Along its collisionless
evolution, a stellar system can be trapped in one of these stationary
solutions and remain frozen in that state for a very long time, until
collisional effects come into play. Physically, the convergence to a
steady state is due to phase mixing and incomplete violent relaxation
\cite{lb,houches,super}. These steady states, referred to as
quasi-stationary states or metaequilibrium states, are described by
non-standard (non-Boltzmannian) distributions $f({\bf r},{\bf
v})$. This is not really surprising since the system is collisionless
so that there is no relation with thermodynamics in the usual sense.
Tsallis distributions $f_{q}({\bf r},{\bf v})$ \cite{tsallis} form a
particular class of stationary solutions of the Vlasov equation. In
fact, these distributions are known for a long time in
astrophysics. They are called {\it stellar polytropes} and were
introduced by Plummer \cite{plummer} in 1911 and Eddington
\cite{edd} in 1916. Although these distributions have interesting
mathematical properties and have been used to construct simple
theoretical models of galaxies \cite{bt}, it is important to note that
they do not provide a good fit of observed galaxies. Once again, this
is not really surprising: there is no fundamental reason why
polytropic distributions (or Tsallis distributions) should be selected
among other stationary solutions of the Vlasov equation
\cite{aa3,gfp}.  Any distribution function $f({\bf r},{\bf
v})=f(\epsilon)$, where $\epsilon={v^{2}/ 2}+\Phi({\bf r})$ is the
stellar energy, is a stationary solution of the Vlasov equation. It
extremizes a functional of the form $S[f]=-\int C(f)d^{3}{\bf r}d^{3}{\bf
v}$, where $C$ is convex, at fixed mass $M$ and energy $E$. Such
functionals have been called H-functions by Tremaine, H\'enon \&
Lynden-Bell \cite{thlb}. Tsallis functional $S_{q}[f]$ is the H-function
associated with stellar polytropes
\cite{aa3}. Furthermore, the {maximization} of a H-function at fixed
mass and energy provides a criterion of nonlinear dynamical stability
for the Vlasov equation \cite{ipser,thlb,aa3,rein}. This ``dynamical
interpretation'' defended by \cite{y} and \cite{aa3,cvb} who view
metaequilibrium states as stationary solutions of the Vlasov equation
and interpret the maximization of a $H$-function as a condition of
nonlinear dynamical stability is different from the ``generalized
thermodynamical interpretation'' defended by
\cite{pp,boghosian,ts,lss,silva,jiulin,rev1,rev2,latora} who view
Tsallis functional $S_{q}[f]$ as a generalized entropy, interpret
quasi-stationary states as generalized maximum entropy states and
consider the maximization of $S_{q}[f]$ as a condition of generalized
thermodynamical stability \footnote{In the context of Vlasov systems,
there is a distinction to make between Tsallis functional
$S_{q}[f]=-{1\over q-1}\int (f^{q}-f)d{\bf r}d{\bf v}$ defined in
terms of the distribution function $f({\bf r},{\bf v})$ and Tsallis
entropy $S_{q}[\rho]=-{1\over q-1}\int (\rho^{q}-\rho)d\eta d{\bf
r}d{\bf v}$ defined in terms of the probability density $\rho({\bf
r},{\bf v},\eta)$ of phase levels. Ignoring the specificities of the
collisionless evolution (Casimir constraints), the authors of
\cite{pp,boghosian,ts,lss,silva,jiulin,rev1,rev2,latora} interprete
$S_{q}[f]$ as a generalized entropy while the correct form of Tsallis
entropy in the context of violent relaxation is $S_{q}[\rho]$ (it
returns the Lynden-Bell entropy for $q=1$ which is the appropriate
form of Boltzmann entropy in that context). By contrast,
\cite{aa3,cvb} interprete the Tsallis functional $S_{q}[f]$ as a $H$-function. In this paper, we only consider the functional $S_{q}[f]$. The important distinction between the  functionals $S_{q}[f]$ and  $S_{q}[\rho]$ is further discussed in \cite{super}.}.

There are two points that we would like to emphasize in this
paper. The first point is that Tsallis (and Boltzmann) distributions
do not have a fundamental importance in regard with the collisionless
dynamics of stellar systems (when the $N\rightarrow +\infty$ limit is
taken before the $t\rightarrow +\infty$ limit).  These are just {\it
particular} stationary solutions of the Vlasov equation
\cite{aa3,gfp}. This is different from the Boltzmann distribution in
regard with the collisional dynamics of stellar systems (when the
$t\rightarrow +\infty$ limit is taken before the $N\rightarrow
+\infty$ limit).  The Boltzmann distribution is of fundamental
importance as it describes the statistical equilibrium state of the
system (up to all the limitations mentioned in the first paragraph).
Furthermore, it is the {\it unique} stationary solution of the kinetic
Landau equation which describes the collisional relaxation of stellar
systems to order $1/N$
\cite{new}. Thus, the first order correction to the Vlasov limit
selects the Boltzmann distribution. The second point that we want to
emphasize is that, formally, the criterion of nonlinear dynamical
stability for collisionless stellar systems described by the
Vlasov-Poisson system {\it resembles} a criterion of generalized
thermodynamical stability in the microcanonical ensemble. Therefore,
we can develop an {\it effective thermodynamical formalism} (E.T.F) to
investigate the nonlinear {dynamical} stability of collisionless
stellar systems \cite{aa3,gfp}. In this {\it thermodynamical analogy},
the notions of entropy, free energy and temperature are effective.  We
stress that our E.T.F. has a meaning different from the one given by
\cite{pp,boghosian,ts,lss,silva,jiulin,rev1,rev2,latora}. 
In our point of view, Tsallis functional $S_{q}[f]$ is related to
dynamics, not thermodynamics, in the case of collisionless stellar
systems. What we do essentially is to use the language of
thermodynamics to study nonlinear dynamical stability problems. We
stress that the ``thermodynamical analogy'' that we develop is only
valid for distribution functions of the form $f=f(\epsilon)$ with
$f'(\epsilon)<0$ that describe a sub-class of {\it spherical} stellar
systems \cite{bt}.  We have also found that, formally, the criterion
of nonlinear dynamical stability for barotropic stars described by the
Euler-Poisson system resembles a criterion of thermodynamical
stability in the canonical ensemble
\cite{aa3}. Since ``canonical stability  implies
microcanonical stability in thermodynamics'', we deduce, using our
thermodynamical analogy, that ``the stability of a barotropic star
implies the stability of the corresponding stellar system (but not the
converse)''. This provides an original interpretation of the nonlinear
Antonov first law in terms of ensemble inequivalence
\cite{aa3}.

Tsallis distributions can also arise in other domains of physics and
astrophysics for different reasons. For example, polytropic
distributions correspond to the unique stationary state of a certain
class of generalized kinetic equations or generalized Fokker-Planck
equations satisfying a H-theorem for the Tsallis entropy (or free
energy) \cite{ppkin,limaprl,cspoly}. However, it is not clear at the
present time to which physical systems these generalized kinetic
equations apply. We have argued that they may be ``effective
equations'' trying to take into account ``hidden constraints'' in
complex systems \cite{gfp}. We have also shown that it is possible to
formally generalize kinetic and Fokker-Planck equations, so that they
increase an arbitrary entropy \cite{gfp}, not necessarily the
Boltzmann or the Tsallis one. In a different context, Taruya \&
Sakagami \cite{tsprl} have shown that the collisional evolution of
stellar systems, between the phase of violent relaxation and the phase
of gravothermal catastrophe, can be described by a sequence of
polytropic distributions with a time varying index, interpreted as
{\it quasi-equilibrium states}. In conclusion, Tsallis distributions
(and other non-standard distributions) arise in different domains of
physics (biology, economy,...) for reasons that are not necessarily
related to thermodynamics.  The important point is to explain, in each
specific situation, what is the {\it physical meaning} of such
distributions. We shall give here various examples in astrophysics but
the concepts that we discuss can apply to other systems with
long-range interactions.

\section{Barotropic stars}
\label{sec_gas}

\subsection{The Euler-Poisson system}
\label{sec_ep}

We consider a self-gravitating gaseous medium described by the Euler-Poisson
system
\begin{equation}
{\partial\rho\over\partial t}+\nabla \cdot (\rho {\bf u})=0,
\label{ep1}
\end{equation}
\begin{equation}
{\partial {\bf u}\over\partial t}+({\bf u}\cdot \nabla)
{\bf u}=-{1\over\rho}\nabla p-\nabla\Phi,
\label{ep2}
\end{equation}
\begin{equation}
\Delta\Phi=4\pi G\rho.
\label{ep3}
\end{equation}
We assume that the gas is barotropic with an equation of state
$p=p(\rho)$.  This provides the simplest model of stars \cite{chandra}.
The total energy of the star is
\begin{equation}
{\cal W}[\rho,{\bf u}]=\int\rho\int_{0}^{\rho}{p(\rho')\over\rho^{'2}}\,
d\rho'd^{3}{\bf r}+{1\over 2}\int \rho\Phi\, d^{3}{\bf r}+\int \rho
{{\bf u}^{2}\over 2}\,d^{3}{\bf r}.
\label{ep4}
\end{equation}
The first term is the work $-p(\rho)d(1/\rho)$ done in compressing it
from infinite dilution, the second term is the gravitational energy
and the third term is the kinetic energy associated with the mean
motion. It is straightforward to verify that the energy functional
(\ref{ep4}) is conserved by the Euler-Poisson system ($\dot {\cal
W}=0$). Therefore, a minimum of ${\cal W}$ at fixed mass determines a
stationary solution of the Euler-Poisson system which is nonlinearly
dynamically stable \cite{holm}. Physically, this means that a small
perturbation remains close (in some norm) to the minimum. We are led
therefore to consider the minimization problem
\begin{eqnarray}
{\rm Min}\ \lbrace {\cal W}[\rho,{\bf u}]\quad |\ M[\rho]=M\rbrace .
\label{ej4}
\end{eqnarray}
This minimization problem is consistent with the phenomenology of
star formation. A star forms through the gravitational collapse of a
molecular cloud until it has reached a state of minimum energy.
Canceling the first order variations of Eq.~(\ref{ep4}), we obtain
${\bf u}={\bf 0}$ and the condition of hydrostatic equilibrium
\begin{equation}
\nabla p=-\rho\nabla\Phi,
\label{ep5}
\end{equation}
between pressure and gravity. Therefore, extrema of ${\cal W}$
correspond to stationary solutions of the Euler-Poisson system
(\ref{ep1})-(\ref{ep3}). Combining the condition of
hydrostatic equilibrium (\ref{ep5}) and the equation of state
$p=p(\rho)$, we get
\begin{eqnarray}
\int^{\rho}{p'(\rho')\over\rho'}\,d\rho'=-\Phi,
\label{ep6}
\end{eqnarray}
so that $\rho$ is a function of $\Phi$ that we note
$\rho=\rho(\Phi)$. Considering now the second order variations, the
condition of nonlinear dynamical stability is
\begin{eqnarray}
\delta^{2}{\cal W}={1\over 2}\int \delta\rho\delta\Phi\, d^{3}{\bf r}+
\int {p'(\rho)\over 2\rho}(\delta\rho)^{2}\,d^{3}{\bf r}\ge 0,
\label{ep7}
\end{eqnarray}
for all perturbations that conserve mass, i.e. $\int \delta\rho\,
d^{3}{\bf r}=0$. We note that the second integral in Eq.~(\ref{ep7})
can be written in a more conventional form by using
$p'(\rho)/\rho=-1/\rho'(\Phi)$.  The minimization problem
(\ref{ej4}) has been studied in detail in \cite{aa1,aa2,aa3} for
spherically symmetric systems described by an isothermal and a
polytropic equation of state (the system can be self-confined or
confined by a box or by an external pressure). It is shown that the
point in the series of equilibria where the system ceases to be a
minimum of ${\cal W}$ and becomes a saddle point coincides with the
point of neutral linear stability.  Therefore, the conditions of
linear and nonlinear dynamical stability coincide.

\subsection{Isothermal and polytropic stars}
\label{sec_lte}

In the theory of stellar structure, it is usually assumed (at least in
the simple models that we consider here) that a star is a perfect gas in
local thermodynamic equilibrium (L.T.E).  Therefore, at each point,
the distribution function of the particles is of the form
\begin{equation}
f({\bf r},{\bf v})=\biggl \lbrack {m\over 2\pi k_{B}T({\bf r})}
\biggr\rbrack^{3/2}\rho({\bf r})e^{-{m \lbrack {\bf v}-{\bf u}({\bf r})
\rbrack^{2}\over 2 k_{B}T({\bf r})}}.
\label{lte1}
\end{equation}
If we define the pressure by $p={1\over 3}\int f w^{2}d^{3}{\bf v}$
where ${\bf w}={\bf v}-{\bf u}({\bf r})$, we find that the local
equation of state is
\begin{equation}
p({\bf r})={\rho({\bf r})\over m}k_{B}T({\bf r}).
\label{lte2}
\end{equation}
We shall now consider two types of models that have been widely
studied in astrophysics, namely  the case of isothermal stars and
the case of polytropic stars.

{\it (i) Isothermal stars:} If we assume that the star is in thermal
equilibrium, then the temperature is uniform: $T({\bf r})=T$. From one
place to the other the transformation is {\it isothermal}. In that
case, the equation of state becomes
\begin{equation}
p({\bf r})={\rho({\bf r})\over m} k_{B}T.
\label{ip0}
\end{equation}
Since the pressure depends only on the density, the fluid  is
barotropic. The energy functional (\ref{ep4}) reads
\begin{equation}
{\cal
W}[\rho,{\bf u}]=k_{B}T\int {\rho\over m}\ln {\rho\over m}\, d^{3}{\bf r}
+{1\over 2}\int \rho\Phi\, d^{3}{\bf r}+\int \rho {{\bf u}^{2}\over
2}\,d^{3}{\bf r}.
\label{ip1}
\end{equation}
We note that the energy functional (\ref{ip1}) of an isothermal
gas coincides with the Boltzmann free energy $F_{B}=E-TS_{B}$ of a
self-gravitating system in the canonical ensemble (see \cite{aa1}
for more details). Furthermore, the distribution function of the
particles is the Maxwell-Boltzmann distribution (with a uniform
temperature). The relation (\ref{ep6}) between the density and the
potential can be explicitly written
\begin{equation}
\rho=A'e^{-{m\Phi\over k_{B}T}}.
\label{ip2}
\end{equation}
According to Eq.~(\ref{ej4}), it can be obtained by extremizing
Eq.~(\ref{ip1}) at fixed mass or by combining the condition of
hydrostatic equilibrium with the isothermal equation of state.

{\it (ii) Polytropic stars:} Stars are not isothermal in
general. In the convective region, they are rather in a situation
where the entropy is uniform: $s({\bf r})=s$. From one
place to the other the transformation is {\it isentropic}. In that
case, using the first principle of thermodynamics $du=-pdv+Tds$ with
$ds=0$ and $du=c_{v}dT$, we obtain the differential equation
$c_{v}dT+pdv=0$ (with $v=m/\rho$). Using the equation of state of a
perfect gas $p={\rho\over m}k_{B} T$ and the Mayer relation
$c_{p}-c_{v}=k_{B}$, the differential equation can be integrated and
yields  the polytropic equation of state
\begin{equation}
p({\bf r})=K\rho({\bf r})^{\gamma},
\label{ip0b}
\end{equation}
where $K$ is the polytropic constant and $\gamma=c_{p}/c_{v}$ is the
ratio of specific heats at constant pressure and constant volume
\cite{chandra}. The polytropic index $n$ is defined by
$\gamma=1+1/n$. The polytropic constant can be written
$K=k_{B}\Theta_{\gamma}/m$, where $\Theta_{\gamma}$ is sometimes
called a polytropic temperature (cf. \cite{chandra}, p. 86). For
$n\rightarrow +\infty$, we recover the isothermal case with
$\gamma=1$ and $\Theta_{\gamma}=T$.  For a polytropic gas, the local
temperature $T({\bf r})$, defined by (\ref{lte2}), is given by
\begin{equation}
{k_{B}T({\bf r})\over m}=K\rho({\bf r})^{1/n}.
\label{ip6}
\end{equation}
We note that the temperature is position dependent (while the specific
entropy $s$ is uniform) and related to the density by a power law;
this is the local version of the usual isentropic law
$TV^{\gamma-1}=Cst.$ in standard thermodynamics. The polytropic index
$n$ is related to the gradients of temperature and density according
to
\begin{equation}
{\nabla\ln T}={1\over n}\nabla\ln\rho.
\label{ip7}
\end{equation}
The energy functional (\ref{ep4}) can be
written
\begin{equation}
{\cal
W}[\rho,{\bf u}]={K\over \gamma-1}\int (\rho^{\gamma}-\rho)\,d^{3}{\bf r}
+{1\over 2}\int \rho\Phi \,d^{3}{\bf r}+\int \rho {{\bf u}^{2}\over
2}\,d^{3}{\bf r}.
\label{ip4}
\end{equation}
For an isentropic transformation ($ds=0$), the first term in
Eqs.~(\ref{ep4}) and (\ref{ip4}) represents the internal energy
since $du/m=-pd(1/\rho)$. For the polytropic equation of state
(\ref{ip0b}), the internal energy is a power-law functional:
$U=K/(\gamma-1)\int \rho^{\gamma} d^{3}{\bf r}$. We have then added
a constant term $K/(\gamma-1)\int \rho d^{3}{\bf r}$, proportional
to the total mass, in the polytropic energy functional (\ref{ip4})
so as to recover the isothermal energy functional (\ref{ip1}) for
$\gamma\rightarrow 1$ (i.e., $n\rightarrow +\infty$). Using
Eq.~(\ref{ep6}) or extremizing the energy functional (\ref{ip4}) at
fixed mass, we find that the relation between the density and the
potential is
\begin{eqnarray}
\rho=\biggl\lbrack \lambda-{\gamma-1\over K\gamma}\Phi\biggr
\rbrack^{1\over \gamma-1}.
\label{ip5}
\end{eqnarray}
We see at this stage some resemblances with Tsallis distributions
and free energies, see Eqs.~(\ref{ip5}) and (\ref{ip4}). However,
there is {\it no} generalized thermodynamics in the problem. The
physics of gaseous stars can be understood in terms of
conventional thermodynamics once realized that adiabatic
transformations are more relevant than isothermal transformations
in stellar structure. We shall see that these formal resemblances
are essentially fortuitous or the mark of a {\it thermodynamical
analogy}.

\section{Stellar systems}
\label{sec_stellar}

\subsection{The statistical equilibrium state}
\label{sec_vh}

For $t\rightarrow +\infty$, a self-gravitating system is expected
to achieve a statistical equilibrium state. Assuming that the
accessible microstates are equiprobable, the most probable
macroscopic distribution function maximizes the Boltzmann entropy
\begin{equation}
S_{B}[f]=-\int {f\over m}\ln {f\over m}\,d^{3}{\bf r}d^{3}{\bf v},
\label{coll1}
\end{equation}
at fixed mass and energy. The Boltzmann entropy can be obtained by a
standard combinatorial analysis. The criterion of thermodynamical stability
is therefore
\begin{equation}
\label{coll2} {\rm Max}\ \lbrace S_{B}[f]\quad |\ E[f]=E, M[f]=M \rbrace.
\end{equation}
It determines the most probable distribution of stars at statistical
equilibrium. Introducing Lagrange multipliers and writing $\delta
S-\beta\delta E-\alpha\delta M=0$, we get the Boltzmann distribution
\begin{equation}
\label{coll3} f=Ae^{-\beta m\epsilon}.
\end{equation}
We note that the statistical equilibrium state depends only on the
energy $\epsilon=v^{2}/2+\Phi({\bf r})$ of the stars (for a
non-rotating system). The collisional evolution is usually described
by the kinetic Landau equation, see e.g. \cite{new}, that we write
symbolically as
\begin{equation}
{\partial f\over\partial t}+{\bf v}\cdot {\partial f\over\partial {\bf
r}}+{\bf F}\cdot {\partial f\over\partial {\bf
v}}=Q_{Landau}(f),\label{landau}
\end{equation}
where ${\bf F}=-\nabla\Phi$ is the self-consistent gravitational field
produced by the stars and $Q_{Landau}$ is the Landau collision
term. The Landau-Poisson system conserves mass and energy and
increases the Boltzmann entropy (H-theorem).  The Boltzmann
distribution is the only stationary solution of this equation
\footnote{As discussed in the Introduction,
the convergence to the Boltzmann distribution function is hampered by
the escape of stars and the gravothermal catastrophe. Even if we
consider idealized situations of box confined systems with energy
larger than the Antonov threshold so that local entropy maxima exist
in theory, the dynamical convergence to these states is not clearly
understood.  Elaborate kinetic theories of self-gravitating systems
taking into account memory effects and spatial delocalization
\cite{kandrup} do not admit a H-theorem and do not show the
convergence toward the Boltzmann distribution (nor to any other simple
distribution). Related problems are also encountered in the kinetic theory
of point vortices
\cite{pvkin}.}.  From the kinetic theory of stellar systems, we can
show that the collisional relaxation time (Chandrasekhar's time)
scales as $t_{relax}\sim (N/\ln N)t_{D}$.

\subsection{The Vlasov-Poisson system}
\label{sec_vp}

For $t\ll (N/\ln N)t_{D}$, we can ignore the collision term in the
kinetic equation. In this regime, a stellar system is described by
the Vlasov-Poisson system
\begin{equation}
{\partial f\over\partial t}+{\bf v}\cdot {\partial f\over\partial {\bf r}}
+{\bf F}\cdot {\partial f\over\partial {\bf v}}=0,\label{vh0}
\end{equation}
\begin{equation}
\Delta\Phi=4\pi G\int f \,d^{3}{\bf v}.\label{vh1}
\end{equation}
The Vlasov description assumes that the evolution of the system is
encounterless. This is a very good approximation for the dynamics of
stars in elliptical galaxies ($N\sim 10^{12}$). For smaller groups of
stars, such as globular clusters ($N\sim 10^{6}$), encounters have to
be taken into account and the dynamics is governed by the
Landau-Poisson system or by the orbit averaged Fokker-Planck equation
\cite{bt}. Here, we exclusively focus on the collisionless
dynamics. {It is thus clear from the beginning that thermodynamics, in
its usual sense, is irrelevant to the present context; it makes sense
only on much longer timescales when stellar encounters have to be
taken into account.}  The Vlasov equation admits an infinite number of
stationary solutions \cite{bt}. During its collisionless evolution, a
stellar system can be trapped in a stable stationary state and remains
frozen in that quasi-stationary state for a very long time (until
collisions come into play and drive the slow relaxation). According to
the Jeans theorem, the general form of stationary solutions of the
Vlasov equation is a function $f=f(I_{1},...,I_{n})$ of the integrals
of motion. Of course, only {\it stable} stationary solutions are of
physical interest. In the general case, it is difficult to derive a
criterion of dynamical stability. We shall restrict ourselves in this
paper to stationary distribution functions of the form $f=f(\epsilon)$
which depend only on the energy of the stars $\epsilon={v^{2}\over
2}+\Phi({\bf r})$. Such distribution functions describe a sub-class of
{\it spherical stellar systems} (in the general case, $f$ depends on
energy $\epsilon$ and angular momentum ${\bf j}={\bf r}{\times} {\bf
v}$) \cite{bt}. For such distributions, it is possible to derive a
general criterion of nonlinear dynamical stability.

We consider the class of functionals
\begin{equation}
S[f]=-\int C(f)\,d^{3}{\bf r}d^{3}{\bf v},\label{vh2}
\end{equation}
where $C$ is an arbitrary convex function. These functionals are
called H-functions \cite{thlb} because they share some analogies with
the Boltzmann H-function in kinetic theory (see the Conclusion)
\footnote{Strictly speaking, the $H$-functions are defined with the
coarse-grained distribution function $\overline{f}({\bf r},{\bf
v},t)$. We shall argue in the Conclusion and in \cite{super} that, in
case of violent relaxation, it is the {\it coarse-grained}
distribution that is a nonlinearly dynamically stable stationary
solution of the Vlasov equation. Therefore, in all the paper, $f$ must
be regarded in fact as the coarse-grained distribution function
$\overline{f}$. Naturally, the stability results also apply to a
steady distribution function $f$ (fine-grained), but this supposes
that the system has been initially prepared in such a stationary
solution of the Vlasov equation, which is relatively
artificial.}. Since these functionals are particular Casimirs, they
are conserved by the Vlasov equation. The total energy $E={1\over
2}\int f v^{2}d^{3}{\bf r}d^{3}{\bf v}+{1\over 2}\int
\rho\Phi d^{3}{\bf r}$ and the total mass $M[f]=\int fd^{3}{\bf
r}d^{3}{\bf v}$ are also conserved by the Vlasov equation. Therefore,
a maximum of $S$ at fixed mass $M$ and energy $E$ determines a
stationary solution $f({\bf r},{\bf v})$ of the Vlasov equation that
is nonlinearly dynamically stable. We are led therefore to consider
the maximization problem
\begin{equation}
\label{vh3} {\rm Max}\ \lbrace S[f]\quad |\ E[f]=E, M[f]=M \rbrace.
\end{equation}
We note that $F[f]=E[f]-TS[f]$, where $T$ is an arbitrary positive constant, is
also conserved by the Vlasov equation (this is called an
energy-Casimir functional \cite{holm}). Therefore, a minimum of $F$ at
fixed mass $M$ is also a nonlinearly dynamically stable stationary
solution of the Vlasov equation. This criterion can be written
\begin{equation}
\label{vh4} {\rm Min}\ \lbrace  F[f]\quad |\ M[f]=M \rbrace,
\end{equation}
where
\begin{equation}
\label{vh5} F[f]={1\over 2}\int f
v^{2}\,d^{3}{\bf r}d^{3}{\bf v}+{1\over 2}\int \rho\Phi \,d^{3}{\bf r}
+T\int C(f)d^{3}{\bf r}\,d^{3}{\bf v}.
\end{equation}
Similar criteria of nonlinear dynamical stability have been introduced
in the context of two-dimensional hydrodynamics described by the 2D
Euler-Poisson system \cite{ellis}. The optimization problem
(\ref{vh4}) corresponds to the nonlinear dynamical stability
criterion of Holm {\it et al.} \cite{holm} and the optimization
problem (\ref{vh3}) corresponds to the refined dynamical stability
criterion of Ellis {\it et al.} \cite{ellis}. These criteria are in
general not equivalent for systems with long-range interactions, and
this is similar to a notion of ensemble inequivalence in
thermodynamics (see Sec. \ref{sec_ta}).  Inspired by these studies, we
have introduced the same criteria to study the nonlinear dynamical
stability of stellar systems described by the Vlasov-Poisson system
\cite{aa3}. Recently, we found that similar optimization principles
have also been introduced independently by Rein and Guo in the
mathematical literature, see
\cite{rein}.

Introducing Lagrange multipliers and writing $\delta S-\beta\delta
E-\alpha\delta M=0$ or $\delta F-\alpha\delta M=0$, we find that the
critical points of the variational problems (\ref{vh3}) and
(\ref{vh4}) are both given by
\begin{equation}
\label{vh6}
C'(f)=-\beta\epsilon-\alpha.
\end{equation}
Since $C'$ is a monotonically increasing function of $f$, we can
inverse this relation to obtain $f=F(\beta\epsilon+\alpha)$ where
$F(x)=(C')^{-1}(-x)$. Therefore, the optimization problems (\ref{vh3})
and (\ref{vh4}) determine stationary solutions of the Vlasov equation
of the form $f=f(\epsilon)$. As said previously, this is a particular
case of the Jeans theorem describing a sub-class of spherical stellar
systems \cite{bt}. From the identity $f'(\epsilon)=-\beta/C''(f)$
resulting from Eq.~(\ref{vh6}), $f(\epsilon)$ is a monotonic
function. Assuming that $f$ is decreasing, which is the physical
situation, imposes $\beta=1/T>0$. Setting $J=S-\beta E$, the condition
of nonlinear dynamical stability provided by the criterion (\ref{vh3})
is
\begin{eqnarray}
\delta^{2}J=-\int C''(f){(\delta f)^{2}\over 2}\,d^{3}{\bf r}d^{3}{\bf v}-
{1\over 2}\beta\int \delta\rho\delta\Phi \,d^{3}{\bf r}\le 0,\nonumber\\
\forall \delta f \quad |\quad  \delta E=\delta M=0, \qquad\qquad\qquad
\label{vh7}
\end{eqnarray}
and the condition of nonlinear dynamical stability provided by the
criterion (\ref{vh4}) is
\begin{eqnarray}
\delta^{2}J=-\int C''(f){(\delta f)^{2}\over 2}\,d^{3}{\bf r}d^{3}{\bf v}-
{1\over 2}\beta\int \delta\rho\delta\Phi \,d^{3}{\bf r}\le 0,\nonumber\\
\forall \delta f \quad |\quad  \delta M=0.\qquad\qquad\qquad
\label{vh8}
\end{eqnarray}
The first integral can be written in a more conventional form by using
$f'(\epsilon)=-\beta/C''(f)$. We note the importance of the first
order constraints in the criteria (\ref{vh7}) and (\ref{vh8}). If
condition (\ref{vh8}) is satisfied for all perturbations that conserve
mass, it is a fortiori satisfied for perturbations that conserve mass
{\it and} energy. Therefore, condition (\ref{vh8}) implies condition
(\ref{vh7}), but not the opposite. We conclude that the stability
criterion (\ref{vh3}) is more refined than the stability criterion
(\ref{vh4}): if a collisionless stellar system satisfies (\ref{vh3})
or (\ref{vh4}), it is nonlinearly dynamically stable; however, if it
does not satisfy (\ref{vh4}), it can be nonlinearly dynamically stable
provided that it satisfies (\ref{vh3}).  This means that we can
``miss'' stable solutions if we use just the optimization problem
(\ref{vh4}). The problem (\ref{vh3}) is richer and allows to construct
a larger class of nonlinearly dynamically stable models. In
particular, complete stellar polytropes with index $3\le n\le 5$
satisfy the criterion (\ref{vh3}) but not the criterion (\ref{vh4})
\cite{aa3}.

\subsection{Thermodynamical analogy}
\label{sec_ta}

To study the nonlinear dynamical stability of a collisionless stellar
system described by a distribution function of the form
$f=f(\epsilon)$ with $f'(\epsilon)<0$, we are led to consider the
optimization problems (\ref{vh3}) and (\ref{vh4}). We note that,
formally, they are {\it similar} to criteria of thermodynamical
stability for collisional self-gravitating systems, see
Eq.~(\ref{coll2}), but they involve a more general functional
$S[f]=-\int C(f)\,d^{3}{\bf r}d^{3}{\bf v}$ (H-function) than the
Boltzmann entropy (\ref{coll1}).  Furthermore, they have a completely
different interpretation as they provide criteria of nonlinear
dynamical stability for a stationary solution of the Vlasov equation
(representing a quasi-stationary structure in the collisionless
regime) while the maximization of the Boltzmann entropy at fixed mass
and energy provides a condition of thermodynamical stability for the
statistical equilibrium state reached for $t\rightarrow +\infty$
(representing the most probable state that we may expect).

\begin{figure}
\centerline{
\psfig{figure=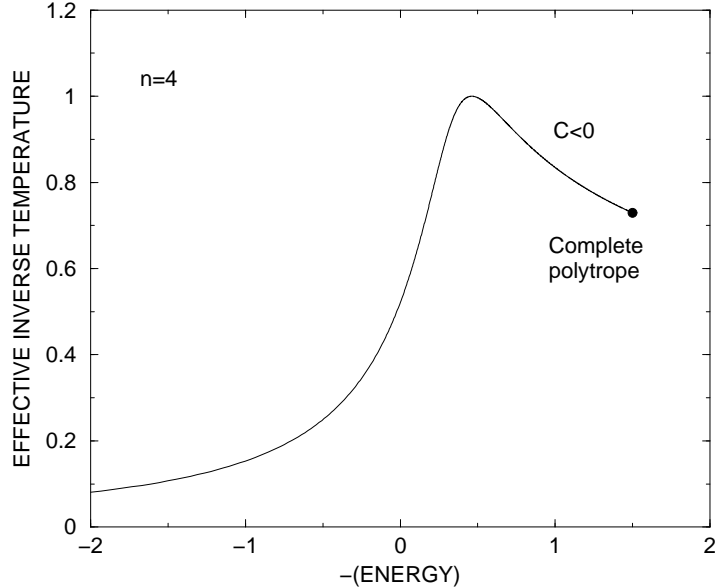,angle=0,height=8cm}}
\caption{Series of equilibria for polytropic spheres
with index $n=4$ \cite{aa2,aa3,cspoly}. In the thermodynamical
analogy, this can be viewed as an effective caloric curve giving the
effective inverse temperature $\eta=(M/4\pi)(4\pi
G/K(n+1))^{n/(n-1)}/R^{(n-3)/(n-1)}$ [which is a monotonic function of
the Lagrange multiplier $\beta$] as a function of energy
$\Lambda=-ER/GM^{2}$. Complete polytropes (with compact support) are
indicated by $(\bullet)$. Since they lie after the turning point of
``temperature'' and before the turning point of energy, in the
thermodynamical analogy, they are unstable in the ``canonical''
ensemble (saddle points of ``free energy'' $F$ at fixed $M$) but
stable in the ``microcanonical'' ensemble (maxima of ``entropy'' $S$
at fixed $E$,$M$). Due to the correspondence that we have found (see
Sec. \ref{sec_anto}), these effective thermodynamical stability
criteria mean in reality that stellar polytropes with $n=4$ are
nonlinearly dynamically stable stationary solutions of the
Vlasov-Poisson system but polytropic stars with $n=4$ are linearly
dynamically unstable stationary solutions of the Euler-Poisson
system. This is similar to a situation of ``ensemble inequivalence''
in thermodynamics. In particular, complete polytropes with $n=4$ lie
in a region of effective negative specific heats $C<0$.}
\label{el4}
\end{figure}

Due to the formal resemblance between the nonlinear dynamical
stability criteria and the thermodynamical stability criteria, we can
develop a {\it thermodynamical analogy} \cite{gfp,aa3} to investigate
the nonlinear dynamical stability of collisionless stellar systems
with $f=f(\epsilon)$ and $f'(\epsilon)<0$. In this analogy, $S$ plays
the role of an ``effective entropy'', $T=1/\beta$ plays the role of an
``effective temperature'' and $F$ plays the role of an ``effective
free energy'' (it is related to $S$ by a Legendre transform). The
criterion (\ref{vh3}) is similar to a condition of ``microcanonical
stability'' and the criterion (\ref{vh4}) is similar to a condition of
``canonical stability''. Since canonical stability implies
microcanonical stability (but not the converse), we recover the fact
that condition (\ref{vh8}) implies condition (\ref{vh7}). The
stability problems (\ref{vh3}) and (\ref{vh4}) can be studied by
plotting the linear series of equilibria $\beta(E)$, which is similar
to a caloric curve in thermodynamics.  According to Poincar\'e's
theorem, a mode of stability is lost or gained at a turning point or
at a branching point. Linear series of equilibria have been studied in
detail in relation with the thermodynamics of self-gravitating systems
for the Boltzmann
\cite{lbw,katz} and Fermi-Dirac \cite{pt} entropies. Due to the above
thermodynamical analogy, the method of linear series of equilibria can
be used similarly to settle the nonlinear dynamical stability of a
collisionless stellar system for a general $H$-function. Note that the
slope $dE/dT$ (where $\beta=1/T$ is the Lagrange multiplier associated
with the energy) plays an important role because, according to the
turning point argument, its change of sign is the signal of an
instability. In thermodynamics, $C=dE/dT$ represents the specific
heat.  Isothermal self-gravitating spheres cease to be minima of free
energy when $dE/dT$ becomes negative passing by $C=+\infty$, and they
cease to be maxima of entropy when $dE/dT$ becomes positive again
passing by $C=0$ \cite{katz}\cite{aa1}. Therefore, the criteria of canonical and
microcanonical stability are inequivalent in the region of negative
specific heats $C<0$. Due to the thermodynamical analogy, we have
similar results for the nonlinear dynamical stability of collisionless
stellar systems (for an arbitrary H-function) but with a different
interpretation. For example, the series of equilibria for polytropic
distributions has been studied in \cite{aa2,aa3,cspoly}. In this
context, the sign of the slope $dE/dT$, which is similar to an effective
 specific heat, is connected to the nonlinear dynamical 
stability of the system (see
Fig. \ref{el4}). {\it This ``dynamical interpretation''
\cite{aa2,aa3,cspoly} is different from the ``generalized thermodynamical
interpretation'' of Taruya \& Sakagami \cite{ts} who considered the
same mathematical problem in relation with Tsallis generalized
thermodynamics.}

\subsection{A new interpretation of the Antonov first law}
\label{sec_anto}

For any spherical stellar system with $f=f(\epsilon)$, there exists a
corresponding barotropic star with the same equilibrium density
distribution. Indeed, defining the density and the pressure by
$\rho=\int fd^{3}{\bf v}=\rho(\Phi)$, $p={1\over 3}\int
fv^{2}d^{3}{\bf v}=p(\Phi)$, and eliminating the potential $\Phi$
between these two expressions, we find that $p=p(\rho)$. Writing
explicitly the density and the pressure in the form
$\rho=4\pi\int_{\Phi}^{+\infty}f(\epsilon)\sqrt{2(\epsilon-\Phi)}d\epsilon$
and $p={4\pi\over 3}\int_{\Phi}^{+\infty}f(\epsilon)\lbrack
2(\epsilon-\Phi)\rbrack^{3/2}d\epsilon$, and taking the derivative of
the second equation, we obtain the condition of hydrostatic
equilibrium (\ref{ep5}). As discussed in Sec. \ref{sec_vh}, the
minimization principle (\ref{vh4}) provides a {\it sufficient}
condition of nonlinear dynamical stability for a collisionless stellar
system. To solve this minimization principle, we can first minimize
$F[f]$ at fixed density $\rho({\bf r})$ to obtain $f_{*}({\bf r},{\bf
v})$. We can then express the functional $F[f]$ as a functional of the
density $\rho({\bf r})$ by setting $F[\rho]=F[f_{*}]$. The
calculations are detailed in \cite{aa3} and the functional $F[\rho]$
can be finally written
\begin{eqnarray}
F[\rho]={1\over 2}\int\rho\Phi\,
d^{3}{\bf r}+\int\rho\int_{0}^{\rho}{p(\rho')\over
\rho'^{2}}\,d\rho'd^{3}{\bf r}, \label{anto4}
\end{eqnarray}
where $p(\rho)$ is the equation of state of the corresponding
barotropic gas. We are now led to consider the minimization problem
\begin{eqnarray}
{\rm Min}\ \lbrace  F[\rho]\quad |\  M[\rho]=M \rbrace .
\label{anto3}
\end{eqnarray}
Now, we observe that the functional (\ref{anto4}) coincides with the
energy functional (\ref{ep4}) of a barotropic gas with ${\bf u}={\bf
0}$. We arrive therefore at the following conclusion: for a given
stellar system with $f=f(\epsilon)$ and $f'(\epsilon)<0$, if we know
that the corresponding barotropic star is nonlinearly dynamically
stable with respect to the Euler-Poisson system, then it is a minimum
of ${\cal W}[\rho,{\bf u}]$, hence of $F[\rho]$ (at fixed
mass). Therefore, the stellar system satisfies the criterion
(\ref{vh4}) so it is nonlinearly dynamically stable with respect to
the Vlasov-Poisson system. This leads to a nonlinear generalization of
the Antonov first law: ``a stellar system is nonlinearly dynamically
stable whenever the corresponding barotropic gas is nonlinearly
dynamically stable'' \cite{aa3}. However, the reciprocal is wrong in
general. A stellar system can be nonlinearly dynamically stable
according to the ``microcanonical'' criterion (\ref{vh3}) while it
does not satisfy the ``canonical'' criterion (\ref{vh4}) so that the
corresponding barotropic star is dynamically unstable \cite{aa2}. Due
to the thermodynamical analogy, the Antonov first law for
collisionless stellar systems has the same status as the fact that:
``canonical stability implies microcanonical stability in
thermodynamics''. The reciprocal is wrong if the ensembles are not
equivalent. To the point of view of their dynamical stability, the
crucial difference between stars and galaxies is that, for galaxies,
both $S$ (H-function) and $E$ are {\it individually} conserved by the
Vlasov equation while in the case of barotropic stars only the total
energy ${\cal W}$ is conserved by the Euler equations. Therefore, the
most refined stability criterion for galaxies is (\ref{vh3}) while the
most refined stability criterion for stars is (\ref{ej4}), which is
included in (\ref{vh3}). This is the intrinsic reason why a spherical
galaxy can be stable even if the corresponding barotropic star is
unstable.

It is interesting to consider the application of these results to
polytropes (see Fig. \ref{el4}). By plotting the series of
equilibria $\beta(E)$, it can be shown that a complete polytrope
(i.e., for which the density drops to zero at a finite radius) ceases
to be a minimum of $F$ for $n> 3$
\cite{aa2}. This is when the slope $dE/dT$ becomes negative in the
series of equilibria. In the thermodynamical analogy, this is similar
to a loss of canonical stability. Using the above result, we conclude
that polytropic stars are nonlinearly dynamically stable for $n< 3$
and they become dynamically unstable for $n> 3$
\cite{aa2}. However, complete polytropes cease to be a maximum of $S$
only for $n< 5$ (they also cease to be self-confined at this index)
\cite{aa3,cspoly}. This is similar to a loss of microcanonical
stability in the thermodynamical analogy. Therefore, stellar
polytropes are nonlinearly dynamically stable for $n< 5$ and they
become dynamically unstable for $n> 5$. For $3<n< 5$, stellar
polytropes are nonlinearly dynamically stable while corresponding
polytropic stars are dynamically unstable. This is similar to a
situation of ensemble inequivalence (in a region of negative specific
heats) in thermodynamics \cite{touchette,bb}. Of course, the dynamical
stability of stellar polytropes and polytropic stars has been
established for a long time in astrophysics \cite{bt} but the
interpretation of these results (and of the Antonov first law) in
terms of a thermodynamical analogy and a situation of ensemble
inequivalence is new.  Furthermore, only linear dynamical stability is
considered in \cite{bt} while the criteria (\ref{vh3}) and (\ref{vh4})
are criteria of {\it nonlinear} dynamical stability. Similar criteria
of nonlinear dynamical stability can be developed for the HMF model
\cite{y,cvb} (in that case, the ``ensembles'' are equivalent) and for
other systems described by Vlasov or Euler equations \cite{new}.

\subsection{Isothermal stellar systems}
\label{sec_vb}

We consider the $H$-function
\begin{eqnarray}
S=-\int f\ln f\, d^{3}{\bf r}d^{3}{\bf v}.
\label{vb1}
\end{eqnarray}
This functional resembles the Boltzmann entropy (\ref{coll1}) in
thermodynamics. However, as explained above, its physical
interpretation is different. Its maximization at fixed mass and
energy determines a nonlinearly dynamically stable stationary
solution of the Vlasov equation with distribution function
\begin{eqnarray}
f=A e^{-\beta \epsilon}.
\label{vb2}
\end{eqnarray}
This distribution function has the same form (but a different
interpretation) as the statistical equilibrium state (\ref{coll3})
of a stellar system resulting from a collisional relaxation or as
the distribution function (\ref{lte1}) of the gas in an isothermal
star. For these reasons, collisionless stellar systems described by
Eq.~(\ref{vb2}) are sometimes called isothermal stellar systems and
$T=\beta^{-1}$ is sometimes called a ``temperature''.  In fact, it
is more proper to regard $T$ as the velocity dispersion of the
stellar system. We note also that the mass $m$ of the individual
particles does not appear in the H-function (\ref{vb1}) and in the
distribution function (\ref{vb2}) contrary to the case of
statistical equilibrium, see (\ref{coll1}) and (\ref{coll3}). This
is because the present description is based on the Vlasov equation
in which the mass of the particles does not appear. Therefore, the
distribution function (\ref{vb2}) does not lead to a segregation by
mass contrary to the statistical equilibrium distribution of a
multi-species self-gravitating system \cite{super}. The barotropic
gas associated with the stellar system defined by Eq.~(\ref{vb2}) is
the isothermal gas with an equation of state $p({\bf r})=\rho({\bf
r}) T$ where $T=1/\beta$. The functional (\ref{anto4}) takes the
form
\begin{eqnarray}
F[\rho]={1\over 2}\int \rho\Phi\, d^{3}{\bf r}+T\int \rho\ln\rho
\,d^{3}{\bf r}.
\label{vb4}
\end{eqnarray}
It coincides with the energy functional (\ref{ip1}) of an isothermal
star with ${\bf u}={\bf 0}$. The density is related to the potential
according to
\begin{eqnarray}
\rho=A' e^{-\beta \Phi}.
\label{vb3}
\end{eqnarray}
This relation can be directly obtained by extremizing $F[\rho]$ at
fixed mass.  We can express the distribution function in terms of the
density according to
\begin{eqnarray}
f=\biggl ({\beta\over 2\pi}\biggr )^{3/2}\rho({\bf r})
\ e^{-\beta {v^{2}\over 2}}.
\label{vb5}
\end{eqnarray}

\subsection{Stellar polytropes}
\label{sec_vb2}

We consider the $H$-function
\begin{eqnarray}
S=-{1\over q-1}\int (f^{q}-f)\,d^{3}{\bf r}d^{3}{\bf v}.
\label{vb6}
\end{eqnarray}
It has the same form as the Tsallis entropy in non-extensive
thermodynamics \cite{tsallis}. However, as explained above, its
physical interpretation is different. In the present context, its
maximization at fixed mass and energy determines a nonlinearly
dynamically stable stationary solution of the Vlasov equation, called
a stellar polytrope, with distribution function
\begin{eqnarray}
f=\biggl \lbrack \mu-{\beta (q-1)\over q}
\epsilon\biggr \rbrack^{1\over q-1}.
\label{vb7}
\end{eqnarray}
The polytropic index is related to the parameter $q$ by
\begin{eqnarray}
n={3\over 2}+{1\over q-1}.
\label{vb7bis}
\end{eqnarray}
Stellar polytropes have been introduced by Plummer \cite{plummer} in
1911. In fact, Plummer did not write the distribution function
explicitly and analyzed the distribution of stars in globular
clusters by treating (incorrectly) the system as a gas in convective
equilibrium. Plummer's law corresponds to the analytical solution of
a polytropic gas with index $n=5$ found by Schuster in 1883.
Eddington \cite{edd} in 1916 criticized the ``adiabatic'' model of
Plummer and justified Plummer's law as a particular solution of the
Vlasov equation satisfying the Jeans theorem. Eddington was
apparently the first to write the distribution function of a
``stellar polytrope'' correctly. Note that the ``mistake'' of
Plummer is interesting because it reflects the mathematical property
that ``for any spherical stellar system with $f=f(\epsilon)$ there
exists (formally) a corresponding barotropic star'' \cite{bt}. Later
on, the fact that polytropic distributions (\ref{vb7}) maximize a
certain Casimir functional at fixed mass and energy, and that this
maximization problem is related to their dynamical stability with
respect to the Vlasov equation has been recognized by several
authors, e.g. Ipser \cite{ipser} in 1974. However, in the work
\cite{ipser} the functional is simply written as $W=-\int
f^{1+1/(n-3/2)}d^{3}{\bf r}d^{3}{\bf v}$. Therefore, when
$n\rightarrow +\infty$, the connection with the functional
(\ref{vb1}) describing isothermal stellar systems is not direct. The
functional (\ref{vb6}) introduced by Tsallis \cite{tsallis} (albeit
in a quite different context) is better to make this connection
because, using L'H\^opital's rule, Eq.~(\ref{vb6}) reduces to
Eq.~(\ref{vb1}) for $q\rightarrow 1$ (or $n\rightarrow +\infty$).
Similarly, the distribution function written in the form (\ref{vb7})
reduces to the distribution function (\ref{vb2}) for $q\rightarrow
1$ (or $n\rightarrow +\infty$). The form given by Eddington
\cite{edd} and by Binney \& Tremaine \cite{bt} is less convenient to
make this connection. {\it Therefore, in the context of Vlasov
systems, we interpret Tsallis functional as a useful H-function
connecting continuously stellar polytropes and isothermal stellar
systems.}

It is shown in \cite{cspoly} that physical polytropic distribution
functions have $q>0$ and
$\beta>0$. We need therefore to consider two cases according to the
sign of $q-1$. For $q>1$ ($n>3/2$), the distribution function can be
written
\begin{eqnarray}
f=A(\alpha-\epsilon)^{1\over q-1},
\label{vb8}
\end{eqnarray}
where we have set $A=\lbrack\beta(q-1)/q\rbrack^{1\over q-1}$ and
$\alpha=q\mu/\beta(q-1)$. It is valid for $v<v_{max}({\bf
r})=\sqrt{2(\alpha-\Phi({\bf r}))}$. For $v>v_{max}({\bf r})$, we
set $f=0$. This situation corresponds to the usual polytropes
considered in astrophysics \cite{bt}. For $q\rightarrow 1$
($n\rightarrow +\infty$), we recover isothermal distributions and
for $n=3/2$ the distribution function is a step function. This is
the distribution function of a Fermi gas at zero temperature
describing classical white dwarf stars \cite{chandra}. Note that
stellar polytropes with $1/2<n<3/2$ exist mathematically but the
distribution function $f(\epsilon)$ increases with the energy so
they may not be physical. The density and the pressure can be
expressed as
\begin{eqnarray}
\rho=4\pi\sqrt{2}A(\alpha-\Phi)^{n}{\Gamma(3/2)\Gamma(n-1/2)\over
\Gamma(n+1)},
\label{vb9}
\end{eqnarray}
\begin{eqnarray}
p={1\over
n+1}4\pi\sqrt{2}A(\alpha-\Phi)^{n+1}{\Gamma(3/2)\Gamma(n-1/2)\over
\Gamma(n+1)}. \label{vb10}
\end{eqnarray}

For $0<q<1$, the distribution function can be written
\begin{eqnarray}
f=A(\alpha+\epsilon)^{1\over q-1},
\label{vb11}
\end{eqnarray}
where we have set $A=\lbrack\beta(1-q)/q\rbrack^{1\over q-1}$ and
$\alpha=q\mu/\beta(1-q)$. In that case, $v_{max}\rightarrow
+\infty$. We shall only consider distributions for which the density
and the pressure (first and second velocity moments) are defined. This implies
$3/5<q<1$ (i.e., $n<-1$). Then, the density and the pressure can be
expressed as
\begin{eqnarray}
\rho=4\pi\sqrt{2}A(\alpha+\Phi)^{n}{\Gamma(-n)\Gamma(3/2)\over \Gamma(3/2-n)},
\label{vb12}
\end{eqnarray}
\begin{eqnarray}
p=-{1\over n+1}4\pi\sqrt{2}A(\alpha+\Phi)^{n+1}{\Gamma(-n)\Gamma(3/2)\over
\Gamma(3/2-n)}.
\label{vb13}
\end{eqnarray}

Eliminating the gravitational potential between the expressions
(\ref{vb9})-(\ref{vb10}) and (\ref{vb12})-(\ref{vb13}) giving the
density and the pressure, one finds that
\begin{eqnarray}
p=K\rho^{\gamma}, \qquad \gamma=1+{1\over n},
\label{vb14}
\end{eqnarray}
where
\begin{eqnarray}
K={1\over n+1}\biggl \lbrace 4\pi\sqrt{2}A {\Gamma(3/2)
\Gamma(n-1/2)\over \Gamma(n+1)}
\biggr \rbrace^{-1/n}, \quad (n>3/2),
\label{vb15}
\end{eqnarray}
\begin{eqnarray}
K=-{1\over n+1}\biggl \lbrace 4\pi\sqrt{2}A{\Gamma(-n)\Gamma(3/2)
\over\Gamma(3/2-n)}\biggr \rbrace^{-1/n}, \quad (n<-1).
\label{vb16}
\end{eqnarray}
Therefore, a stellar polytrope has the same equation of state as a
polytropic star. {\it However, we stress that they do not have the
same distribution function, except for $n\rightarrow +\infty$
(isothermal system)}. The distribution function of the gas in a
polytropic star is given by the local thermodynamic equilibrium
(\ref{lte1}) with a space dependent temperature (\ref{ip6}) while the
distribution function of a stellar polytrope is given by (\ref{vb7})
which is completely different. The functional (\ref{anto4}) can be put
in the form
\begin{eqnarray}
F[\rho]={1\over 2}\int \rho\Phi \,d^{3}{\bf r}+{K\over\gamma-1}
\int (\rho^{\gamma}-\rho)\, d^{3}{\bf r}.
\label{vb17}
\end{eqnarray}
It coincides with the energy functional (\ref{ip4}) of a polytropic
star with ${\bf u}={\bf 0}$.  Minimizing this functional at fixed
mass, we directly obtain the relation between the density and the
potential
\begin{eqnarray}
\rho=\biggl\lbrack \lambda-{\gamma-1\over K\gamma}\Phi\biggr
\rbrack^{1\over \gamma-1}.
\label{vb18}
\end{eqnarray}
It resembles Tsallis distribution (in physical space) where $\gamma$
plays the role of the $q$-parameter and $K$ plays the role of the
temperature $\beta^{-1}$ (compare Eqs.~(\ref{vb7}) and
(\ref{vb18})). Also, the functional (\ref{vb17}) resembles Tsallis
free energy $F[\rho]=E[\rho]-KS_{\gamma}[\rho]$ in physical space.
{\it In this sense, Tsallis distributions are ``stable laws'' since
they keep a similar structure as we pass from phase space
$f=f(\epsilon)$ to physical space $\rho=\rho(\Phi)$ with the
correspondence $\gamma\leftrightarrow q$ and $K\leftrightarrow
\beta^{-1}$.} This is probably the only class of distributions
enjoying this property. However, we again emphasize that, in the
present context, the resemblance with a thermodynamical formalism is
essentially fortuitous or effective.

We can also express the distribution in terms of the density according to
\begin{eqnarray}
f={1\over Z}\biggl \lbrack \rho({\bf r})^{1/n}-{v^{2}/2\over
(n+1)K}\biggr\rbrack^{n-3/2}, \label{vb19}
\end{eqnarray}
\begin{eqnarray}
Z=4\pi\sqrt{2}{\Gamma(3/2)\Gamma(n-1/2)\over\Gamma(n+1)}
\lbrack K(n+1)\rbrack^{3/2}, \quad (n>3/2)
\label{vb20}
\end{eqnarray}
\begin{eqnarray}
Z=4\pi\sqrt{2}{\Gamma(-n)\Gamma(3/2)\over\Gamma(3/2-n)}\lbrack
-K(n+1)\rbrack^{3/2}, \quad (n<-1). \label{vb21}
\end{eqnarray}
This is the counterpart of the isothermal distribution function
(\ref{vb5}). For any stellar system, we define the {\it kinetic
temperature} by ${3\over 2}T({\bf r})={1\over 2}\langle v^{2}\rangle$
or $p({\bf r})=\rho({\bf r}) T({\bf r})$. It is proportional to the
local velocity dispersion of the particles. For a polytropic
distribution function
\begin{eqnarray}
T({\bf r})=K\rho({\bf r})^{1/n}.
\label{vb21bis}
\end{eqnarray}
As indicated previously, for collisionless stellar systems $T$ must
be regarded as a mean-squared velocity rather than a temperature
(the individual mass of the particles never appears in the
equations). We shall use, however, the temperature terminology.
Using Eq.~(\ref{vb21bis}), the distribution function (\ref{vb19})
can be written
\begin{eqnarray}
f=B_{n}{\rho({\bf r})\over \lbrack 2\pi T({\bf r})\rbrack^{3/2}}
\biggl\lbrack 1-{v^{2}/2\over (n+1)T({\bf r})}\biggr\rbrack^{n-3/2},
\label{vb23}
\end{eqnarray}
\begin{eqnarray}
B_{n}={\Gamma(n+1)\over\Gamma(n-1/2)(n+1)^{3/2}}, \quad (n>3/2)
\label{vb24}
\end{eqnarray}
\begin{eqnarray}
B_{n}={\Gamma(3/2-n)\over\Gamma(-n)\lbrack -(n+1)\rbrack^{3/2}} \quad (n<-1).
\label{vb25}
\end{eqnarray}
Note that for $n>3/2$, the maximum velocity can be expressed
in terms of the kinetic temperature according to
\begin{eqnarray}
v_{max}({\bf r})=\sqrt{2(n+1)T({\bf r})}.
\label{vb26}
\end{eqnarray}
Using $\Gamma(z+a)/\Gamma(z)\sim z^{a}$ for $z\rightarrow +\infty$,
we recover the isothermal distribution (\ref{vb5}) for $n\rightarrow
+\infty$.  On the other hand, from Eqs.~(\ref{vb21bis}) and
(\ref{vb18}), we immediately get $T({\bf
r})=K(\lambda-(\gamma-1)\Phi({\bf r})/K\gamma)$ so that
\begin{eqnarray}
\nabla T=-{\gamma-1\over\gamma}\nabla\Phi.
\label{vb22}
\end{eqnarray}
This relation can also be obtained directly from Eqs.~(\ref{vb8})
and (\ref{vb23}).  This shows that, for a stellar polytrope, the
kinetic temperature is a linear function of the gravitational
potential (this is also true for a polytropic gas). The coefficient
of proportionality $(\gamma-1)/\gamma=1/(n+1)=2(q-1)/(5q-4)$ is
related to the polytropic index. There is an interesting application
of this result related to a remark of Eddington \cite{edd}. First of
all, for isolated stellar systems, the energy $\epsilon_{max}$ at
which the distribution function vanishes should correspond to the
escape energy $\epsilon=0$. This implies $\alpha=0$ in
Eq.~(\ref{vb8}) which leads to the ordinary DF of a polytrope (the
distribution (\ref{vb8}) with $\alpha\neq 0$ is in fact a
generalization of the polytropic model with $\epsilon_{max}\neq 0$).
Now, $\alpha=0$ implies that $\lambda=0$ in Eq.~(\ref{vb18}), see,
e.g., Eq.~(\ref{vb9}). Since $\rho$ and $\Phi$ must then vanish at
the same point, this implies that the density profile must go to
$+\infty$ (with a finite mass) which is only possible for the
polytropic index $n=5$. This is Eddington's interpretation of
Plummer's law \cite{edd}. On the other hand, for $\lambda=0$, we get
$T({\bf r})=-(\gamma-1)\Phi({\bf r})/\gamma$. Defining the local
kinetic energy by $E_{kin}={1\over 2}\rho \langle
v^{2}\rangle={3\over 2}\rho({\bf r}) T({\bf r})$ and the local
potential energy by $E_{pot}={1\over 2}\rho\Phi$, we get
$E_{kin}/E_{pot}=3T({\bf r})/\Phi({\bf r})=-3(\gamma-1)/\gamma$.
Thus, a {\it local} Virial theorem $2E_{kin}+E_{pot}=0$ holds for
$\gamma=6/5$, i.e. $n=5$. Distribution functions satisfying this
relation have been called hypervirial models of stellar systems
\cite{evans}.

\subsection{Generalized temperatures}
\label{sec_temp}

We shall now discuss different notions of ``temperature" that
emerge in the case of polytropic distributions. In the case of
isothermal distribution functions, all these notions coincide.
However, for polytropic distribution functions they have a
different character.

{(i) Lagrange multiplier $\beta=1/T_{0}$}: the Lagrange multiplier
$\beta$ appearing in the variational principle $\delta S-\beta\delta
E-\alpha\delta M=0$ associated with the optimization problem
(\ref{vh3}) can be viewed as a generalized inverse temperature since
$\beta=dS/dE$. In terms of this Lagrange multiplier $\beta$, a
polytropic distribution function can be written
\begin{eqnarray}
f=\biggl \lbrack \mu-{\beta (q-1)\over q}\epsilon\biggr
\rbrack^{1\over q-1}.
\label{temp0}
\end{eqnarray}
As indicated previously, the series of equilibria $\beta(E)$ is
similar to a caloric curve and the slope $C=dE/dT_{0}$ is similar to a
specific heat. The sign of $C$ is connected to the nonlinear dynamical
stability of the system via the turning point argument. This implies
that the correct ``temperature'' to use in the definition of $C$ when
using it to settle the nonlinear dynamical stability of a
collisionless stellar system is the inverse of the Lagrange multiplier
$\beta$. However, $T_{0}=1/\beta$ has generally {\it not} the
dimension of a temperature.

{(ii) Dimensional temperature:} we can define a quantity that has the
dimension of an inverse temperature by setting $b=\beta/\mu(\beta)$
(note that the normalization condition $M=\int f d^{3}{\bf r}d^{3}{\bf
v}$ imposes that $\mu$ is an implicit function of $\beta$). If we
define furthermore $A(b)=\mu(b)^{1/(q-1)}$, we can rewrite the
polytropic distribution function in terms of $b$ as
\begin{eqnarray}
f=A(b)\biggl \lbrack 1-{b (q-1)\over q}\epsilon\biggr
\rbrack^{1\over q-1}. \label{temp0b}
\end{eqnarray}
Therefore, we see that the polytropic distribution satisfies a
relation of the form
\begin{eqnarray}
f=A(b)F(b\epsilon)
\label{temp0c}
\end{eqnarray}
with $F(x)=\lbrack 1-((q-1)/q)x\rbrack^{1/(q-1)}$. Not all
distribution functions of the form $f=F(\beta\epsilon+\alpha(\beta))$,
coming from the optimization problem (\ref{vh3}), can be written as
(\ref{temp0c}). Ponno \cite{ponno} has shown in a different context
that only polytropic distribution functions (Tsallis q-distributions)
satisfy this relation.

{(iii) Polytropic temperature $K$}: we have indicated that the
polytropic constant $K$ shares some analogies with a temperature and,
for this reason, it is sometimes called a polytropic temperature. We
note, in particular, that $K$ is uniform in a polytropic system, as is
the temperature in an isothermal system. The distribution function of
a stellar polytrope can be written in terms of $K$ as
\begin{eqnarray}
f={1\over Z}\biggl \lbrack \rho({\bf r})^{1/n}-{v^{2}/2\over
(n+1)K}\biggr\rbrack^{n-3/2}. \label{temp1}
\end{eqnarray}
We note that, for a stellar polytrope, $K$ is a monotonic function of
the Lagrange multiplier $\beta$ expressed  by Eqs.~(\ref{vb15}) and
(\ref{vb16}) so they essentially play the same role. Therefore, it
is often convenient to use $K$ instead of $\beta$ when we plot the
series of equilibria for polytropes (this is the convention that we
have adopted in \cite{aa2,aa3,cspoly} and in Fig. \ref{el4}).

{(iv) Kinetic temperature:} For any stellar system, it is natural to
introduce a kinetic temperature $T({\bf r})$ through the relation
${3\over 2}T({\bf r})={1\over 2}\langle v^{2}\rangle$. In general, the
kinetic temperature is inhomogeneous. The distribution function of a
stellar polytrope can be written in terms of $T({\bf r})$ as
\begin{eqnarray}
f=B_{n}{\rho({\bf r})\over \lbrack 2\pi T({\bf r})
\rbrack^{3/2}}\biggl\lbrack 1-{v^{2}/2\over (n+1)T({\bf r})}
\biggr\rbrack^{n-3/2},
\label{temp3}
\end{eqnarray}
where $T({\bf r})=p(\rho)/\rho=K\rho({\bf r})^{\gamma-1}$. Combined
with Eq.~(\ref{vb18}), this immediately yields the relation
(\ref{vb22}) between the kinetic temperature and the gravitational
potential.

{(v) Energy excitation temperature:} as indicated in \cite{gfp}, in
any system with $f=f(\epsilon)$, one may define a local energy
dependent excitation temperature by the relation
\begin{eqnarray}
{1\over T(\epsilon)}=-{d\ln f\over d\epsilon}.
\label{svv1}
\end{eqnarray}
For the isothermal distribution (\ref{vb2}), $T(\epsilon)$ coincides
with the temperature $T=1/\beta$. For the polytropic distribution
(\ref{vb7}), $T(\epsilon)=q\mu/\beta-(q-1)\epsilon$. This excitation
temperature has a constant gradient
\begin{eqnarray}
{dT\over d\epsilon}=1-q,
\label{temp4}
\end{eqnarray}
related to Tsallis $q$ parameter (or equivalently to the index $n$ of
the polytrope). The other parameter $\mu$ is related to the value of
energy where the temperature reaches zero. Relation (\ref{temp4}) is
similar to that found by Almeida \cite{almeida} in a different
context.

\section{A second interpretation: generalized kinetic equations}
\label{sec_genkin}

In Sec. \ref{sec_stellar}, using the orthodox interpretation of
astrophysics \cite{bt}, we have justified the polytropic distribution
function (\ref{vb7}) as a particular stationary solution of the Vlasov
equation. This is probably the most relevant justification of this
distribution function in astrophysics since most astrophysical bodies
are governed by the collisionless Boltzmann equation. In this context,
the connection with a thermodynamical formalism is essentially
effective (thermodynamical analogy). The Tsallis functional $S_{q}[f]$
is a H-function whose maximization at fixed mass and energy provides a
criterion of nonlinear dynamical stability. Furthermore, the
polytropic distribution does not play any special role among other
stationary solutions of the Vlasov equation. Its main interest is its
mathematical simplicity since it possesses properties of homology in
continuity with the isothermal distribution \cite{chandra}.

Polytropic distribution functions can also emerge in a different
context.  For example, Silva \& Alcaniz \cite{silva} and Du Jiulin
\cite{jiulin2} assume that the dynamics of their system is governed by
a generalized Boltzmann equation admitting a H-theorem for the Tsallis
entropy \cite{limaprl}. In that case, the polytropic distribution is
the {\it only} stationary solution of this kinetic equation. However,
the relevance of this generalized kinetic equation for astrophysical
systems remains to be established (Silva \& Alcaniz and Du Jiulin do
not mention to which systems their approach applies). Furthermore, as
shown in Chavanis \cite{gfp}, it is possible to construct even more general
kinetic equations that increase a larger class of ``generalized
entropies'' than the one proposed by Tsallis. Consequently, it has to
be explained why Tsallis entropy should be selected among other
functionals of the form $S=-\int C(f)d^{3}{\bf r}d^{3}{\bf
v}$. Concerning the theoretical study of self-gravitating polytropic
spheres, our results differ from those obtained by Silva \& Alcaniz
\cite{silva}. A first difference is that we define the kinetic
temperature by $T=p/\rho=({1\over 3}\int fv^{2}d^{3}{\bf v})/\int f
d^{3}{\bf v}={1\over 3}\langle v^{2}\rangle$ while they use
$q$-expectation values.  Another important difference is that we
consider an inhomogeneous medium while Silva \& Alcaniz consider a
homogeneous system. In particular, they derive a distribution function
of the form
\begin{eqnarray}
f=B_{q}\biggl\lbrack 1-(1-q){m v^{2}\over 2k_{B}T}\biggr\rbrack^{1/(1-q)},
\label{silva}
\end{eqnarray}
instead of Eq.~(\ref{temp3}). Our generalization is important
because gravitational systems are inhomogeneous. Thus, the equation
of state associated with Tsallis distribution is that of a polytrope
$p=K\rho^{\gamma}$ while the approach of Silva \& Alcaniz
\cite{silva} gives the impression that the equation of state is
isothermal with a generalized $q$-temperature: $p={2\over 5-3q}\rho
{k_{B}T\over m}$. This is an artifact of their homogeneity
assumption. In fact, for inhomogeneous self-gravitating systems, the
kinetic temperature associated with a polytropic distribution is
space-dependent, see Eq.~(\ref{vb21bis}). Therefore, the stability
condition given in \cite{silva} based on the sign of the specific
heat is not correct. The stability of polytropic spheres must take
into account spatial inhomogeneity  as discussed in
\cite{ts,cspoly}.

Du Jiulin \cite{jiulin2} considers inhomogeneous distribution
functions of self-gravitating systems described by Tsallis
statistics. However, his approach is unnecessary complicated
because he did not realize at the start that the stationary
distribution function must be a function of the individual energy
$\epsilon={v^{2}\over 2}+\Phi({\bf r})$ alone. Indeed, for the class
of generalized mean-field kinetic equations considered in \cite{gfp},
and written symbolically as
\begin{equation}
{\partial f\over\partial t}+{\bf v}\cdot {\partial f\over\partial {\bf
r}}-\nabla\Phi\cdot {\partial f\over\partial {\bf v}}=
Q_{C}(f),\label{gkin}
\end{equation}
where $C$ is a convex function determining the equilibrium state, a
stationary solution must cancel {\it independently} the advective term
and the collision term. The cancellation of the advective term implies
that $f$ is a stationary solution of the Vlasov equation.  This is
satisfied by any $f=F(\epsilon)$ where $F$ is arbitrary.  Then, $F$ is
determined by the collision term by requiring $Q_C(f)=0$ (see
\cite{gfp} for more details). We stress that the stationary solution
$f_{C}(\epsilon)$ of the kinetic equation (\ref{gkin}) only depends on
the energy $\epsilon$ while the Vlasov equation admits a much larger
family of stationary distributions. This can also be seen by noting
that the kinetic equations in \cite{gfp} increase a generalized
entropy functional $S=-\int C(f)d^{3}{\bf r}d^{3}{\bf v}$ at fixed
mass and energy. Therefore, the stationary state maximizes $S$ at
fixed $E$, $M$ and this implies $f=f_{C}(\epsilon)$. When $S$ is the
Tsallis entropy, this directly leads to Eq.~(\ref{temp0}) given in
Chavanis \cite{gfp} while Du Jiulin postulates the form (\ref{temp3})
as an extension of the homogeneous distribution (\ref{silva}) given in
Silva \& Alcaniz \cite{silva}. He then obtains Eq.~(\ref{vb22}) by a
laborious perturbative calculation while this relation is exact and
results almost immediately from Eq.~(\ref{temp0}). In addition, Du
Jiulin argues that this relation provides an analytic expression of
the $q$-parameter. This is not the logical way of reading this
relation. The relation (\ref{vb22}) is just a property of a polytropic
distribution function with index $q$, not an explanation of the
meaning of Tsallis $q$-parameter.  Equation (\ref{vb22}) is not
mysterious (nor fundamental) once the physical meaning of the
temperature is given. In our approach, $T({\bf r})$ represents the
usual kinetic temperature related to the density and pressure via
$T({\bf r})=p({\bf r})/\rho({\bf r})$. Said differently,
$E_{kin}={1\over 2}\rho \langle v^{2}\rangle={3\over 2}\rho T$ is the
local density of kinetic energy. Since $\rho({\bf r})$ and $p({\bf
r})$ are functions of the gravitational potential $\Phi({\bf r})$, see
Sec. \ref{sec_anto}, this implies that $T({\bf r})$ is a function of
$\Phi({\bf r})$. This is true for any distribution function of the
form $f=f(\epsilon)$. Now, for a polytropic distribution, $T({\bf
r})=p({\bf r})/\rho({\bf r})=K\rho({\bf r})^{\gamma-1}$ where
$\rho(\Phi({\bf r}))$ is explicitly given by Eq.~(\ref{vb18}). This
directly leads to Eq.~(\ref{vb22}) stating that $T$ is a linear
function of $\Phi$ in that case.

\section{A third interpretation: quasi-equilibrium states of
a ``collisional'' dynamics}
\label{sec_ts}

In a recent numerical study, Taruya \& Sakagami \cite{tsprl} find that
the transient stages of the collisional relaxation of the $N$-stars
system confined within a box can be fitted by a sequence of polytropic
distribution functions with a time dependent $q(t)$ parameter. This is
an interesting result which contrasts from the King's sequence
(truncated isothermals) applying to tidally truncated stellar
systems. Therefore, in their latest works \cite{tsprl}, Taruya \&
Sakagami interpret polytropic distributions as quasi-equilibrium
states of a ``collisional'' dynamics. However, this is essentially an
out-of-equilibrium result so that the connection with Tsallis
thermodynamics is not clear. In particular, the evolution of stellar
clusters is usually described by the orbit-averaged-Fokker-Planck
equation which exploits the timescale separation between the dynamical
time and the relaxation time. It is possible that the time dependent
solutions of the orbit-averaged-Fokker-Planck equation can be fitted
by $q(t)$-distributions although this equation is based on standard
statistical mechanics and satisfies a H-theorem for the ordinary
Boltzmann entropy, not for the Tsallis entropy. Therefore, the results
of Taruya \& Sakagami \cite{tsprl} tend to disfavor Tsallis
thermodynamics instead of establishing it: indeed, $q(t)$-polytropes
can emerge from a kinetic equation based on standard thermodynamics
(Boltzmann H-theorem).  In addition, it seems unlikely that
$q(t)$-polytropes are {\it exact} solutions of the
orbit-averaged-Fokker-Planck equation, so that they represent more a
convenient fit for the solutions of a complicated nonlinear time
dependent partial differential equation than a fundamental outcome of
a ``generalized thermodynamics''. Furthermore, it is not clear whether
they attract all the solutions of the orbit-averaged-Fokker-Planck
equation or if their ``basin of attraction'' is limited to some
particular initial conditions.  Finally, it is not clear whether the
interpretation of polytropic distributions functions given by Taruya
\& Sakagami \cite{tsprl} as quasi-equilibrium states of a collisional
dynamics extends to other systems with long-range interactions such as
the HMF model for example. These are important questions to
consider. In any case, the numerical experiments of Taruya \& Sakagami
\cite{tsprl} fill the gap between the description of the ``violent
relaxation'' (first stage) and the description of the ``gravothermal
catastrophe'' (ultimate stage) of stellar systems.

\section{Conclusion}
\label{sec_conclusion}

We have discussed three distinct interpretations of Tsallis functional
$S_{q}[f]$ in astrophysics. Most discussion has been devoted to
Vlasov-Poisson systems.  Systems with long-range interactions
spontaneously form ``coherent structures'' that are not described in
general by the Boltzmann distribution. This is the case for galaxies
in astrophysics, vortices in 2D turbulence and quasi-stationary states
in the HMF model. Two types of interpretation have been given to
describe these structures. In the ``generalized thermodynamical
interpretation''
\cite{pp,boghosian,ts,lss,silva,jiulin,rev1,rev2,latora},
the metaequilibrium states are viewed as generalized statistical
equilibrium states. The maximization of the Tsallis entropy $S_{q}[f]$
at fixed mass and energy provides a condition of generalized
thermodynamical stability.  In the ``dynamical
interpretation'' given here and in \cite{y,aa3,cvb}, the
metaequilibrium states are particular stationary solutions of the
Vlasov equation. Any DF $f=f(\epsilon)$ with $f'(\epsilon)<0$
extremizes a certain functional $S[f]=-\int C(f)d{\bf r}d{\bf v}$,
called a H-function, at fixed mass and energy. The condition of
maximum is a criterion of nonlinear dynamical stability via the Vlasov
equation. These two interpretations are physically different.
However, formally, the criterion of nonlinear dynamical stability for
collisionless stellar systems described by the Vlasov equation
resembles a criterion of thermodynamical stability in the
microcanonical ensemble and the criterion of nonlinear dynamical
stability for barotropic stars described by the Euler equation
resembles a criterion of thermodynamical stability in the canonical
ensemble. We have used this ``thermodynamical analogy'' to provide an
original interpretation of the Antonov first law in terms of ensemble
inequivalence. However, the resemblance with a thermodynamical
formalism is effective.

In this paper, we have not discussed {\it how} a collisionless stellar
system described by the Vlasov-Poisson system can reach a
quasi-stationary distribution (meta-equilibrium) and {what} determines
this metaequilibrium state. This is discussed in \cite{houches,super}
based on the concept of phase mixing and violent
relaxation introduced by Lynden-Bell
\cite{lb} in 1967. As a result of a mixing process, the
coarse-grained distribution function $\overline{f}({\bf r},{\bf v},t)$
achieves a stationary state on a very short timescale (of the order of
a few dynamical times) while the fine-grained distribution function
${f}({\bf r},{\bf v},t)$ develops intermingled filaments at smaller
and smaller scales. Lynden-Bell tried to predict the optimal
metaequilibrium state by using statistical mechanics arguments. This
prediction is complicated because we have to deal with an infinite
number of constraints (Casimirs). Therefore, the coarse-grained
distribution function arising from the statistical theory of violent
relaxation is non-Boltzmannian in general \cite{aa3}. It can be viewed
as a sort of superstatistics \cite{super}. In the dilute limit of his
theory, Lynden-Bell \cite{lb} predicted an isothermal distribution
$\overline{f}\sim e^{-\beta\epsilon}$. He also understood that his
prediction fails for high energies (corresponding to extended orbits)
due to the complicated problem of {\it incomplete relaxation}: the
systems tends to reach the statistical equilibrium state (most mixed)
but cannot attain it because the fluctuations of the gravitational
field, which are the engine of the relaxation, rapidly die away. This
is why other stationary solutions of the Vlasov equation can emerge in
practice corresponding to partially mixed states leading to confined
structures. This kinetic confinement can be understood by developing
dynamical models of violent relaxation \cite{csr,kinvr}.  The
metaequilibrium state reached by the system as a result of incomplete
violent relaxation is very difficult to predict because it depends on
the efficiency of mixing. It is different in general from Tsallis'
distribution (pure polytrope) \cite{gfp}. For example, stellar
polytropes do not provide good models of galaxies \cite{bt}. A better
model is a composite model \cite{hm,aa3} with an isothermal core
($n=\infty$, $q=1$) justified by Lynden-Bell's theory of violent
relaxation and a polytropic halo ($n=4$, $q=7/5$) resulting from
incomplete relaxation. In this context, we have proposed to interpret
the $q$ parameter, which can vary in space, as a {\it measure of mixing}
\cite{aa3}.  These general concepts also apply to 2D vortices in
hydrodynamics \cite{houches} and to the HMF model
\cite{y,cvb}. 

We note that the maximization problem (\ref{vh3}) determining the
nonlinear dynamical stability of collisionless stellar systems is
consistent with the phenomenology of violent relaxation
\cite{lb,thlb,aa3}. Indeed, the H-functions 
$H[\overline{f}]=-\int C(\overline{f}) d^{3}{\bf r}d^{3}{\bf v}$
calculated with the coarse-grained distribution function
$\overline{f}({\bf r},{\bf v},t)$ increase with time ($-H$ decreases),
a property similar to the H-theorem of thermodynamics, while the
energy and the mass remain approximately conserved. However, contrary
to the Boltzmann equation, the Vlasov equation does not single out a
unique H-function and the increase of the H-functions is not
necessarily monotonic. Because of this generalized {\it selective
decay principle} \cite{super}, we may expect the system to achieve a
metaequilibrium state which maximizes one of the H-functions (non
universal) at fixed mass and energy. This is consistent with the
criterion of nonlinear dynamical stability (\ref{vh3}) provided that
we interpret the distribution function as the {\it coarse-grained}
distribution function. It is remarkable that these two maximization
principles are consistent although they have a very different physical
content (there is no notion of ``dissipation'' in the first while this
lies at the heart of the second due to coarse-graining). Plastino \&
Plastino \cite{pp} correctly made the link with H-functions but they
also gave the impression that Tsallis functional is special because of
its thermodynamical properties. In addition, Plastino \& Plastino
\cite{pp} did not explain why we have to {\it maximize} the Tsallis
functional. This is important because, as discussed above, this
maximization problem is related to nonlinear dynamical stability, not
thermodynamical stability. In particular, the
notion of normalized $q$-averages (as done in paper III of \cite{ts})
is not relevant to settle the nonlinear dynamical stability of
collisionless stellar systems.

In conclusion, our discussion shows that the notions of dynamics and
thermodynamics are intermingled. There are lot of analogies (formal or
real) and differences. This is a little bit similar to the
wave/particle duality in quantum mechanics, each aspect having been
defended by different schools. The coherent structures that form in
long-range systems have both dynamical and thermodynamical (mixing)
properties. We think, however, that the correct description of
quasi-equilibrium states is, fundamentally, in terms of nonlinear
dynamical stability arguments. Indeed, a collisionless stellar system
can be trapped in a steady state of the Vlasov equation that does {\it
not} satisfy the ``thermodynamic-looking'' criterion (\ref{vh3}). The
maximization principle (\ref{vh3}) leads to DF of the form
$f=f(\epsilon)$ with $f'(\epsilon)<0$ depending only on energy so that
it only characterizes {\it spherical} stellar systems which seldomly
occur in nature. Now, there exists stationary solutions of the Vlasov
equation, consistent with the Jeans theorem
\cite{bt}, that do not depend on the energy alone.  This is the case
for most observed stellar systems. In that case all the analogies with
thermodynamics break down.  This ``problem'' does not occur for 1D
systems such as the HMF model because $f=f(\epsilon)$ is the general
form of stationary solutions of the Vlasov equation for these
systems.  The case of stellar systems is therefore interesting
with respect to 1D systems to show the limitations of the
``effective'' thermodynamical formalism.  In conclusion, the dynamics
and thermodynamics of systems with long-range interactions is rich and
complex.  We have given a lengthy discussion of these concepts in
\cite{houches,aa3,gfp,super,new,cspoly}. In summary, we come to the following
scenario:

{\it

1. The rapid emergence of coherent structures in stellar dynamics
(galaxies), two-dimensional turbulence (jets and vortices) and in the
HMF model (quasi-stationary states) can be explained by a theory of
{\it incomplete violent relaxation} based on the Vlasov or Euler
equation, see \cite{houches}. Formally, the Vlasov equation is
obtained when the $N\rightarrow +\infty$ limit is taken before the
$t\rightarrow +\infty$ limit.

2. The H-functions $H[\overline{f}]$ calculated with the
coarse-grained distribution function $\overline{f}({\bf r},{\bf v},t)$
increase, while the energy $E[\overline{f}]$ and the mass
$M[\overline{f}]$ are conserved \cite{thlb}.

3. Due to violent relaxation and phase mixing \cite{lb}, the
coarse-grained distribution function $\overline{f}({\bf r},{\bf v},t)$
converges toward a steady state $\overline{f}({\bf r},{\bf v})$ which
is a stationary solution of the Vlasov equation. The convergence takes
a few dynamical times, independent on $N$. The fine-grained
distribution ${f}({\bf r},{\bf v},t)$ develops filaments at smaller
and smaller scales and does not achieve a steady state (presumably).

4. If mixing is complete, $\overline{f}({\bf r},{\bf v})$ is given by
Lynden-Bell's theory of violent relaxation \cite{lb}. It is in general
a non-standard (non Boltzmannian) distribution depending on the
initial conditions through the Casimirs. It can be viewed as a form of
superstatistics \cite{super}.

5. If mixing is incomplete, the Lynden-Bell prediction fails (by
definition!) since $\overline{f}({\bf r},{\bf v})$ is only partially
mixed. Incomplete relaxation \cite{lb} can be explained by dynamical
models of violent relaxation which involve a diffusion coefficient
that vanishes in certain regions of phase-space and for large times
\cite{csr,kinvr}.

6.  Since $\overline{f}({\bf r},{\bf v})$ results from a complicated
mixing process, it is a nonlinearly dynamically stable stationary
(i.e., robust) solution of the Vlasov equation. The general form of
stationary solutions is given by the Jeans theorem \cite{bt}. We
emphasize that, in theories of violent relaxation, we must view the
steady distribution function as the {\it coarse-grained} DF. It is in
general very difficult to predict the state actually reached by the
system in case of incomplete relaxation as it depends on the
efficiency of mixing which itself depends on the ``route to
equilibrium'' (dynamics). It does not appear to be explained by the
generalized thermodynamics introduced by Tsallis in an attempt to
describe non-ergodic behaviours \cite{tsallis,super}.

7. If $\overline{f}({\bf r},{\bf v})=\overline{f}(\epsilon)$ with
$\overline{f}'(\epsilon)<0$, which is a particular case of the Jeans
theorem, it extremizes a H-function at fixed mass and energy
\cite{thlb}. Furthermore, the condition of {\it maximum} is a
condition of nonlinear dynamical stability
\cite{aa3}. In that case, we can develop a {\it thermodynamical
analogy} \cite{aa3,gfp} to investigate the nonlinear dynamical
stability problem, but all the notions of thermodynamics are formal or
effective. Furthermore, this thermodynamical analogy can describe only
spherical stellar systems with $f=f(\epsilon)$ and $f'(\epsilon)<0$.

8. Tsallis functional $S_{q}[\overline{f}]$ is the H-function
corresponding to stellar polytropes which form a {\it particular}
class of spherical collisionless stellar systems that are steady
solutions of the Vlasov equation \cite{aa3,gfp}.

9. For $t\rightarrow +\infty$, the system tends to relax toward a
statistical equilibrium state described by the Boltzmann distribution.
Formally, the collisional statistical equilibrium state is obtained when the
$t\rightarrow +\infty$ limit is taken before the $N\rightarrow
+\infty$ limit. In the case of stellar systems, the convergence to
equilibrium is hampered by the escape of stars and the gravothermal
catastrophe (see the Introduction).

10. For intermediate collision times, the evolution of the system is
described by the orbit averaged Fokker-Planck equation
\cite{bt}. The DF passes by a sequence of quasi-equilibrium states
$f(\epsilon,t)$ which are quasi-stationary solutions of the Vlasov
equation slowly evolving in time due to close encounters (finite $N$
effects). In some cases of box confined systems, this sequence is
well-approximated by stellar polytropes with a time dependent index
\cite{tsprl}.

}

We have also discussed another, completely independent, interpretation
of Tsallis functional in relation with generalized kinetic
equations. In this context, Tsallis functional $S_{q}[f]$ {is} a
generalized entropy taking into account ``hidden constraints'' acting
on the underlying dynamics of the system.  We do not reject the
possibility that this generalized thermodynamical approach can be
relevant to certain self-gravitating systems (not stellar systems) but
we stress that the interpretation of Tsallis functional
(Sec. \ref{sec_genkin}) is then completely different from the one
given for Vlasov systems (Sec. \ref{sec_stellar})
\cite{cspoly}. Furthermore, as shown in
\cite{gfp}, this generalized thermodynamical formalism (and related
kinetic theories) can be extended to a larger class of functionals
than the one proposed by Tsallis.

\end{document}